\documentclass[aps,prl,twocolumn,superscriptaddress,showpacs,byrevtex]{revtex4-1}
\usepackage{etex}
\usepackage{epsfig}
\usepackage{dcolumn}
\usepackage{bm}
\usepackage{color}
\usepackage{lineno}

\def\bd{\begin{document}} \def\ed{\end{document}}
\def\bmp{\begin{minipage}} \def\emp{\end{minipage}}
\def\bcc{\begin{center}} \def\ecc{\end{center}}     \def\npg{\newpage}
\def\beq{\begin{equation}} \def\eeq{\end{equation}} \def\hph{\hphantom}
\def\be{\begin{equation}} \def\ee{\end{equation}} \def\r#1{$^{[#1]}$}
\def\n{\noindent} \def\ni{\noindent} \def\pa{\parindent}
\def\hs{\hskip} \def\vs{\vskip} \def\hf{\hfill} \def\ej{\vfill\eject}
\def\cl{\centerline} \def\ob{\obeylines}  \def\ls{\leftskip}
\def\underbar#1{$\setbox0=\hbox{#1} \dp0=1.5pt \mathsurround=0pt
   \underline{\box0}$}   \def\ub{\underbar}    \def\ul{\underline}
\def\f{\left} \def\g{\right} \def\e{{\rm e}} \def\o{\over} \def\d{{\rm d}}
\def\vf{\varphi} \def\pl{\partial} \def\cov{{\rm cov}} \def\ch{{\rm ch}}
\def\la{\langle} \def\ra{\rangle} \def\EE{e$^+$e$^-$} \def\pt{p_{\rm t}}
\def\bitz{\begin{itemize}} \def\eitz{\end{itemize}}
\def\btbl{\begin{tabular}} \def\etbl{\end{tabular}}
\def\btbb{\begin{tabbing}} \def\etbb{\end{tabbing}}
\def\beqar{\begin{eqnarray}} \def\eeqar{\end{eqnarray}}
\def\\{\hfill\break} \def\dit{\item{-}} \def\i{\item}
\def\bbb{\begin{thebibliography}{9} \itemsep=-1mm}
\def\ebb{\end{thebibliography}} \def\bb{\bibitem}
\def\bpic{\begin{picture}(260,240)} \def\epic{\end{picture}}
\def\akgt{\cl{\bf ACKNOWLEDGMENTS}}
\def\fgn{\noindent{\bf\large\bf figure captions}}
\def\lan{\langle}
\def\ran{\rangle}
\def\p{\pi}
\def\ifmath#1{\relax\ifmmode #1\else $#1$\fi}%
\def\rc{\ifmath{{\mathrm{c}}}}
\def\cut{\ifmath{{\mathrm{cut}}}}
\def\rF{\ifmath{{\mathrm{F}}}}
\def\rK{\ifmath{{\mathrm{K}}}}
\def\rp{\ifmath{{\mathrm{p}}}}
\def\rt{\ifmath{{\mathrm{t}}}}
\def\LAB{\ifmath{{\mathrm{LAB}}}}
\def\cut{\ifmath{{\mathrm{cut}}}}
\def\beq{\begin{equation}}
\def\eeq{\end{equation}}

\newcommand{\cinst}[2]{$^{\mathrm{#1}}$~#2\par}
\newcommand{\crefi}[1]{$^{\mathrm{#1}}$}
\newcommand{\crefii}[2]{$^{\mathrm{#1,#2}}$}
\newcommand{\crefiii}[3]{$^{\mathrm{#1,#2,#3}}$}
\newcommand{\HRule}{\rule{0.5\linewidth}{0.5mm}}

\bd

\title{Observation of $Z_c(3900)^{0}$ in $e^+e^-\to\pi^0\pi^0 J/\psi$}
\author{
{\small M.~Ablikim$^{1}$, M.~N.~Achasov$^{9,f}$, X.~C.~Ai$^{1}$, O.~Albayrak$^{5}$, M.~Albrecht$^{4}$, D.~J.~Ambrose$^{44}$, A.~Amoroso$^{48A,48C}$, F.~F.~An$^{1}$, Q.~An$^{45,a}$, J.~Z.~Bai$^{1}$
% ...
\\
\vspace{0.2cm}
(BESIII Collaboration)\\
\vspace{0.2cm}
{\it
$^{1}$ Institute of High Energy Physics, Beijing 100049, People's Republic of China\\
% ...
$^{4}$ Bochum Ruhr-University, D-44780 Bochum, Germany\\
$^{5}$ Carnegie Mellon University, Pittsburgh, Pennsylvania 15213, USA\\
% ...
$^{9}$ G.I. Budker Institute of Nuclear Physics SB RAS (BINP), Novosibirsk 630090, Russia\\
% ...
$^{44}$ University of Rochester, Rochester, New York 14627, USA\\
$^{45}$ University of Science and Technology of China, Hefei 230026, People's Republic of China\\
% ...
$^{48}$ (A)University of Turin, I-10125, Turin, Italy; (B)University of Eastern Piedmont, I-15121, Alessandria, Italy; (C)INFN, I-10125, Turin, Italy\\
% ...
$^{a}$ Also at State Key Laboratory of Particle Detection and Electronics, Beijing 100049, Hefei 230026, People's Republic of China\\
% ...
$^{f}$ Also at the Novosibirsk State University, Novosibirsk, 630090, Russia\\
% ...
}}
\vspace{0.4cm}
}


\begin{abstract}
Using a data sample collected with the BESIII detector operating at the BEPCII storage ring, we observe a new neutral state $Z_c(3900)^{0}$ with a significance of $10.4\sigma$. The mass and width are measured to be $3894.8\pm2.3\pm3.2$ MeV/$c^2$ and $29.6\pm8.2\pm8.2$~MeV, respectively, where the first error is statistical and the second systematic. The Born cross section for $e^+e^-\to\pi^0\pi^0 J/\psi$ and the fraction
of it attributable to $\pi^0 Z_c(3900)^{0}\to\pi^0\pi^0 J/\psi$ in the range $E_{cm}=4.19-4.42$ GeV are also determined.  We interpret this state as the neutral partner of the four-quark candidate $Z_c(3900)^\pm$. 

\end{abstract}


\maketitle

A new charged charmonium-like particle $Z_c(3900)^{\pm}$ has recently been observed through its decay to $\pi^\pm J/\psi$ by BESIII, Belle and a Northwestern University group using CLEO-c data~\cite{Ablikim:2013mio, Liu:2013dau, Xiao:2013iha}.  This state lies just above the threshold for $D\overline{D}^*$ production, similar to the bottomonium-like resonances $Z_b(10610)^\pm$ and $Z_b(10650)^\pm$ that have been observed by Belle at an energy just above $B\overline{B}^*$ threshold~\cite{Belle:2011aa}.  BESIII also observed a structure, $Z_c(3885)^\pm$, in the process $e^+e^-\to\pi^{\pm}(D\overline{D}^*)^{\mp}$, with mass close to $Z_c(3900)^{\pm}$\cite{Ablikim:2013xfr}. Because the $Z_c^{\pm}$ couples to charmonium and has electric charge, it can not be a conventional $q \bar{q}$ meson, but must contain at least two light quarks in addition to a $c \bar{c}$ pair.  Proposed interpretations for $Z_c^{\pm}$ include hadronic molecules, hadro-quarkonia, tetraquark states, and kinematic effects \cite{models}.  The precise structures of the $Z_c^{\pm}$ and other ``$XYZ$'' states remains unknown, and hence that their further study will lead to a deeper understanding of the strong interaction in the non-perturbative regime.  

Progress in clarifying this picture requires measurements of improved precision and searches for additional states.  The first definitive observation of a neutral $Z_c$ state was a BESIII measurement of $Z_c(4020)^0 \to \pi^0 h_c$~\cite{Ablikim:2014dxl}.  Previously, 3.5$\sigma$ evidence for a candidate state $Z_c(3900)^{0}$ decaying to $\pi^0J/\psi$ was observed in report~\cite{Xiao:2013iha}.  In this Letter, we report the observation of $Z_c(3900)^0$ in the process $e^+e^-\to\pi^0\pi^0 J/\psi$ based on data collected with the BESIII detector at the BEPCII electron-positron collider.  First measurements of the Born cross section for $e^+e^-\to\pi^0\pi^0 J/\psi$ and of the fraction of $\pi^0\pi^0 J/\psi$ production attributable to $Z_c(3900)^{0}$ as a function of center-of-mass energy (${E_{\rm cm}}$) are also presented.   Our data sample has an integrated luminosity of $2809.4$~pb$^{-1}$ distributed over the ${E_{\rm cm}}$ range from 4.190 to 4.420~GeV~\cite{ref:ecm}, with an overall measurement uncertainty of $1.0\%$.  The three largest samples have ${E_{\rm cm}}=$~4.230~GeV (1091.7~pb$^{-1}$), 4.260~GeV (825.7~pb$^{-1}$) and 4.360~GeV (539.8~pb$^{-1}$), with the remainder distributed comparably among seven other energies~\cite{Ablikim:2015nan}.

BESIII is a general-purpose magnetic spectrometer ~\cite{ref:bes3} with a helium-gas-based drift chamber~(MDC), a plastic scintillator time-of-flight system~(TOF), and a CsI(Tl) Electromagnetic Calorimeter~(EMC) enclosed in a superconducting solenoidal magnet providing a 1.0 T field.  The solenoid is supported by an octagonal flux-return yoke with resistive-plate counters interleaved with steel for muon identification (MUC).  


To study the signal response in the BESIII detector, we use a Monte Carlo (MC) package based on GEANT4~\cite{Agostinelli:2002hh} to produce simulated samples for $e^+e^-\to\pi^0 Z_c^0$, $Z_c^0\to\pi^0 J/\psi$ and $e^+e^-\to\pi^0\pi^0 J/\psi$ without an intermediate $Z_c^0$, in both cases with $J/\psi \to e^+e^-$ or $\mu^+\mu^-$. We generate $e^+e^-\to\pi^0 Z_c^0$ and $Z_c^0\to\pi^0 J/\psi$ with isotropic angular distributions. We simulate $e^+e^-\to \pi^0\pi^0 J/\psi$ with a generator of final states with a $J/\psi$ and two pseudoscalars in EVTGEN~\cite{Lange:2001uf,ref:bes3gen} and no intermediate resonances contributing to the $\pi^0$ $\pi^0$ production.  To determine the $Z_c^0$ mass resolution, a signal sample is generated at  $E_{cm}=4.260$~GeV with a $Z_c^0$ mass of 3.9~GeV/$c^2$ and zero width.  In measuring the cross section $\sigma(e^+e^-\to\pi^0\pi^0 J/\psi)$ and ratio 
\begin{eqnarray}
 R=\frac{\sigma(e^+e^-\to\pi^0 Z_c(3900)^{0}\to\pi^0\pi^0 J/\psi)}{\sigma(e^+e^-\to\pi^0\pi^0 J/\psi)}, 
  \end{eqnarray}
  
\noindent MC samples for $e^+e^-\to\pi^0\pi^0 J/\psi$, with and without an intermediate $Z_c^0$, and using mass and width values obtained in this analysis, are generated at all ten ${E_{\rm cm}}$ points.  QED radiative corrections for $J/\psi\to \ell^+\ell^-$ are incorporated with {\sc photos}~\cite{Barberio:1993qi}, and initial-state radiation (ISR) is simulated with KKMC~\cite{ref:kkmc} using the same parameters as in Ref.~\cite{Ablikim:2013mio}.  To study background, a generic $Y(4260)$ sample and a sample of simulated continuum $q\bar{q}$ production at ${E_{\rm cm}}=4.260$~GeV equivalent to 500 pb$^{-1}$ are used, as in Ref.~\cite{Ablikim:2013mio}.

Charged tracks are reconstructed from MDC hits. To optimize the momentum measurement, we restrict the angular range of tracks to be $|\cos\theta|~<~0.93$, where $\theta$ is the polar angle with respect to the positron beam.  We require tracks to pass within $\pm 10$~cm of the interaction point in the beam direction and within $1$~cm in the plane perpendicular to the beam.  Electromagnetic showers are reconstructed by clustering EMC energy deposits.  Efficiency and energy resolution are improved by including energy deposited in nearby TOF counters.  Photons are selected by requiring showers with minimum energies of 25~MeV for $|\cos\theta|<0.8$ or 50~MeV for $0.86 < |\cos\theta| < 0.92$. The angle between the shower direction and the extrapolation of any track to the EMC must be greater than $5^\circ$. A requirement on the EMC timing suppresses electronic noise and deposits unrelated to the event.  Candidates for $\pi^0\to\gamma\gamma$ decays  are selected by requiring the diphoton invariant mass to be in the range
  $100 < M_{\gamma\gamma} < 160$~MeV/$c^2$.

We search for $e^+e^-\to\pi^0\pi^0 J/\psi$ in events with exactly two good oppositely charged tracks and at least four good photons.  In reconstructing $J/\psi \to e^+e^-$, electron candidates must satisfy $E/p>0.7$, where $E$ is the EMC energy and $p$ is the momentum measured in the MDC.  To suppress the small two-photon and Bhabha background, the two-track opening angle is required to be less than $175^\circ$ for any $e^+$ ($e^-$) with $\cos\theta>0.5$ ($\cos\theta<-0.5$).  In selecting $J/\psi \to \mu^+\mu^-$ we require both muon candidates to satisfy $E/p<0.3$ and at least one to have associated hits in more than six MUC layers.    

We reconstruct $\pi^0\pi^0 J/\psi$ candidates if the dilepton invariant mass is within the $J/\psi$ signal region ($2.95<M_{\ell \ell}<3.2$~GeV$/c^2$).  We loop over $\pi^0$ candidates and select the two that do not share photons and have the smallest $\chi^2=\chi^2_{1C}+\chi^2_{\rm 4C}$, where $\chi^2_{1C}$ is the sum of the $\chi^2$ values for the two one-constraint (1C) kinematic fits to the $\pi^0$ mass, and $\chi^2_{\rm 4C}$ is the $\chi^2$ for the 4C fit to the $\pi^0\pi^0 J/\psi$ hypothesis requiring 4-momentum conservation. To suppress combinatorial background we require that there be fewer than two $\pi^0\pi^0$ combinations meeting the tighter $\pi^0$ criterion of   $120 < M_{\gamma\gamma} < 150$~MeV/$c^2$.   

To search for $Z_c^0$ and suppress non-$\pi^0\pi^0 J/\psi$ events, the event is subjected to a 7C fit, adding mass constraints for both $\pi^0$s and the $J/\psi$ to 4-momentum conservation.  To improve resolutions, for events with $\chi^2_{\rm 7C}<$230, the 7C-constrained momenta are used to construct $M_{\pi^0 J/\psi}$ and $M_{\pi^0\pi^0}$. We verified that resonant structures in the $\pi^0\pi^0$ mass spectrum, such as
 $f_0(980)$, do not produce  a peak in the $M_{\pi^0 J/\psi}$ distribution.  Figure~\ref{fig-mpi0jpsi} shows the $\pi^0 J/\psi$ invariant mass distribution in data and the MC-determined background for $E_{cm}=4.260$~GeV.   Each $\pi^0\pi^0J/\psi$ event appears twice, once for each $\pi^0$.  Background processes are estimated by MC to contribute $\sim 12\%$ of selected events, dominated by $XJ/\psi (X\neq\pi^0\pi^0)$ and multi-pion final states. Because the location of the lower peak depends on $E_{cm}$ while the higher peak remains fixed, we interpret the excess near 3.9~GeV/$c^2$ as $Z_c(3900)^{0}$ production and that near 3.4~GeV/$c^2$ as its kinematic reflection.


\begin{figure}[htbp]
\begin{center}
\includegraphics[width=8.5cm]{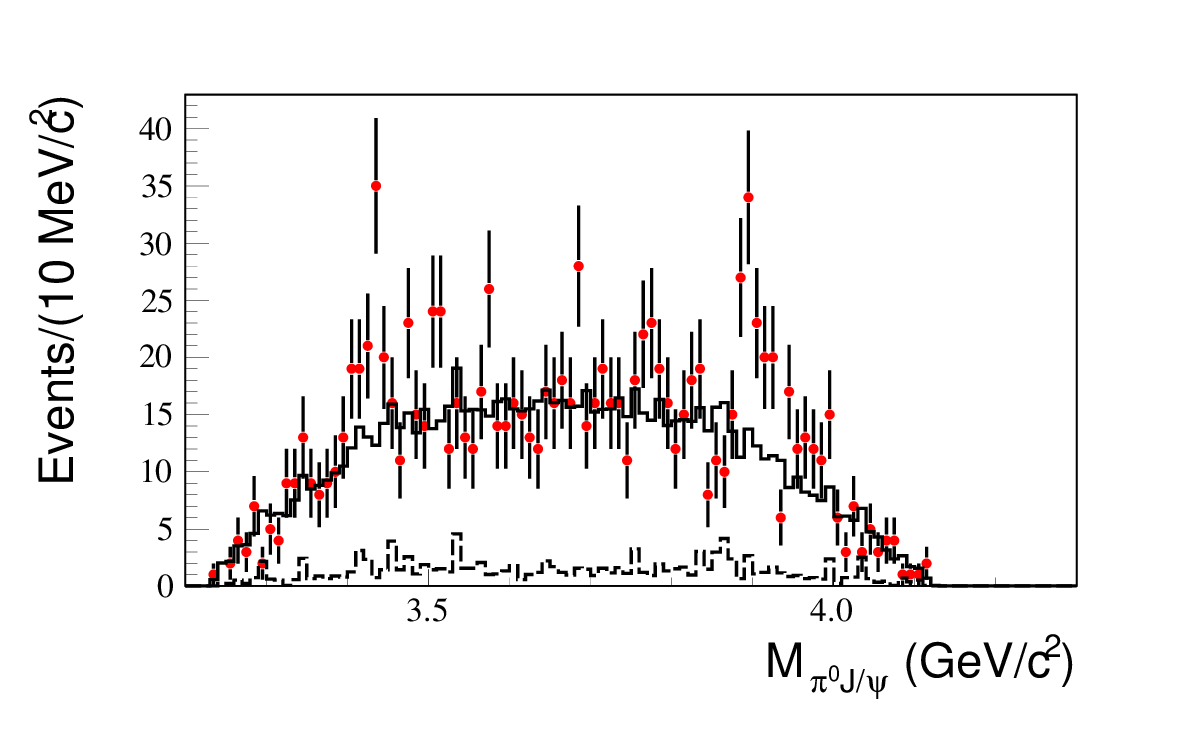}
\caption{Invariant mass distribution for $\pi^0 J/\psi$ candidates in  ${E_{\rm cm}}=4.260$~GeV data (points).  The dashed histogram shows the MC background and the solid histogram is the sum of this background and $\pi^0\pi^0 J/\psi$ production not from $Z_c^0$.} \label{fig-mpi0jpsi}
\end{center}
\end{figure}

%

We extract the yields and resonance parameters of $Z_c(3900)^{0}$ by  performing an unbinned maximum likelihood fit simultaneously to the $\pi^0 J/\psi$ mass distributions for the three high-statistics samples. The fit lower limit is set to 3.65~GeV$/c^2$ to avoid double-counting.  The signal shape is an $S$-wave Breit-Wigner with phase-space factor $pq$, where $p$ is the $Z_c^0$ momentum in the $e^+e^-$ frame and $q$ is the $J/\psi$ momentum in the $Z_c^0$ frame. It is convolved with a resolution function consisting of three Gaussians with parameters set by fitting the zero-width $e^+e^-\to\pi^0 Z_c^0$ MC  sample at ${E_{\rm cm}}=4.260$~GeV (average resolution $\approx 6$~MeV/$c^2$).  The background shape is an ARGUS function~\cite{Albrecht:1990am}. We use the same Breit-Wigner and resolution functions for all energy points because resolution dependence on ${E_{\rm cm}}$ is determined by MC simulation to be very small. The ARGUS parameters are varied independently in the fit, except that the cut-off is based on ${E_{\rm cm}}$. 

Figure~\ref{fig-sim3pzc} shows the simultaneous fit to the three $\pi^{0}J/\psi$ invariant mass distributions, which returns a $Z_c(3900)^{0}$ signal with a statistical significance of 10.4$\sigma$ and a $\chi^2$ of 176 for 151 degrees of freedom.  Yields at ${E_{\rm cm}}=4.230$, 4.260 and 4.360~GeV are 225.3$\pm$41.0, 83.2$\pm$20.5, and 47.5$\pm$12.7, respectively, with a sum of 356.0$\pm$47.6.  The $Z_c(3900)^{0}$ mass and width values with statistical errors are 3894.8$\pm$2.3~MeV/$c^2$ and 29.6$\pm8.2$~MeV, respectively.

\begin{figure}[htbp]
\begin{center}
\includegraphics[height=9cm, width=9cm]{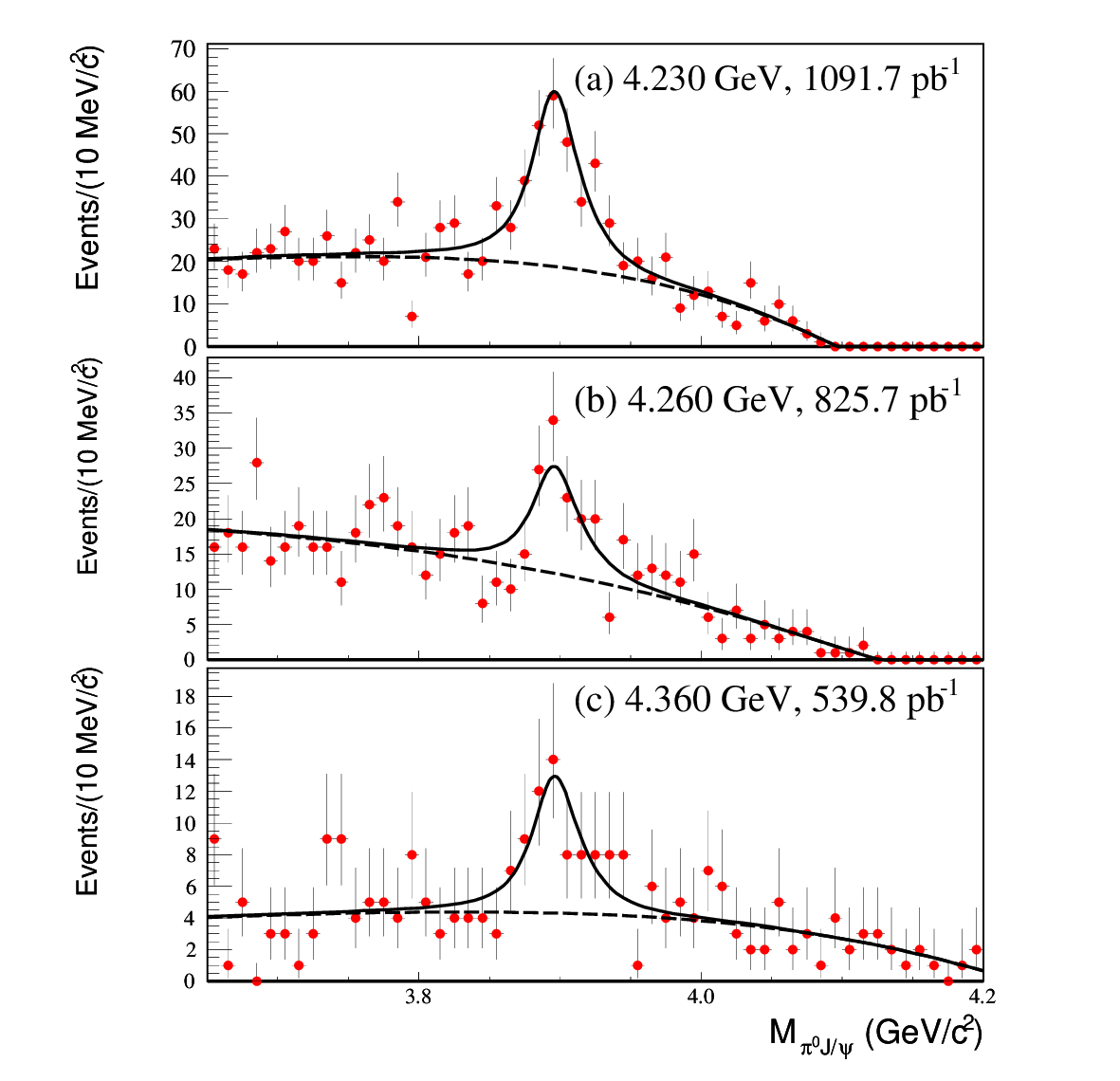}
\caption{The simultaneously fitted $\pi^0 J/\psi$ mass spectra (55 bins in $M_{\pi^0 J/\psi}$) for (a) ${E_{\rm cm}}=4.230$ GeV, (b) ${E_{\rm cm}}=4.260$~GeV, and (c) ${E_{\rm cm}}=4.360$~GeV. Dots represent the data, solid lines represent the fitted results and dashed lines represent fitted backgrounds.} \label{fig-sim3pzc}
\end{center}
\end{figure}

We determine the cross section ratio $R$ and the $e^+e^-\to\pi^0\pi^0 J/\psi$ Born cross section as functions of ${E_{\rm cm}}$ by measuring yields of $Z_c^0$ ($N(Z_c^0)$) and $\pi^0\pi^0 J/\psi$ ($N(\pi^0\pi^0 J/\psi)$).  $N(Z_c^0)$ is determined with a simultaneous fit of the $\pi^0 J/\psi$ mass spectra for all ten ${E_{\rm cm}}$ samples. The signal function is the same as for the fit to the high-statistics samples, with the $Z_c(3900)^{0}$ mass and width fixed to the results of that fit.  Background shapes are ARGUS functions with the cut-off based on ${E_{\rm cm}}$ and other parameters constrained to be the same for all points. 


To obtain $N(\pi^0\pi^0 J/\psi)$, the dilepton mass spectra for all energies are fitted simultaneously.  The small peaking background from $XJ/\psi (X\neq\pi^0\pi^0)$ is treated as a systematic error.  For this determination the 7C kinematic fit including $J/\psi$ mass constraints is inappropriate and the 4C fit results are used.  Events are selected with a cut of $\chi_{\rm 4C}^2 < 80$ based on an optimization considering statistical and systematic uncertainties.  Each signal shape is a Breit-Wigner convolved with a double-Gaussian. The Breit-Wigner is fixed to the width of the $J/\psi$ and the mass is allowed to vary to allow for possible mis-calibration of the momentum scale for reconstructed tracks.  The mean of the first Gaussian of the resolution function is fixed to zero, while the other parameters are varied. The background shape is a first-order Chebyshev polynomial with free parameters. In this fit, the parameters of the double-Gaussian and the polynomial are constrained to be same for all energy points, except for the normalization factor.

The fraction of $\pi^0 \pi^0 J/\psi$ production attributable to $Z_c(3900)^{0}$ is determined with Eq.~\ref{eq:R}, where $\epsilon(Z_c^0)$ is the efficiency for extracting the $Z_c^0$ signal by the fit to the $\pi^0J/\psi$ invariant mass distribution, and $\epsilon_{1}(\pi^0\pi^0 J/\psi)$ and $\epsilon_{2}(\pi^0\pi^0 J/\psi)$ are efficiencies for determining $\pi^0\pi^0 J/\psi$ yields by fits to dilepton mass distributions for processes without and with an intermediate $Z_c^0$, respectively.

\begin{eqnarray}
 R  &=& \frac{N(Z_c^0)}{\epsilon(Z_c^0)}\Big / \Big [ \frac{N(Z_c^0)}{\epsilon(Z_c^0)}+(N(\pi^0\pi^0 J/\psi)- \nonumber \\
   & &\frac{N(Z_c^0)}{\epsilon(Z_c^0)}\epsilon_{2}(\pi^0\pi^0 J/\psi))/\epsilon_{1}(\pi^0\pi^0 J/\psi) \Big ]
  \label{eq:R}
  \end{eqnarray}
The observed cross section for $e^+e^-\to\pi^0\pi^0 J/\psi$ is calculated using Eq.~\ref{eq:obs_CS}, where ${\mathcal L}$ is the integrated luminosity and $\epsilon(\pi^0\pi^0 J/\psi)$ is the weighted average of the efficiencies for events with a $Z_c^0$ ($\epsilon_{2}(\pi^0\pi^0 J/\psi)$) and without a $Z_c^0$ ($\epsilon_{1}(\pi^0\pi^0 J/\psi)$).  The branching ratios ${\cal B}(J/\psi\to e^+e^-)$ and ${\cal B}(J/\psi\to \mu^+\mu^-) $ are taken from the PDG~\cite{ref:PDG_2014}.  

\begin{eqnarray}
 \sigma_{obs}  &=& N(\pi^0\pi^0 J/\psi) \Big / \Big [{\mathcal L}\times \epsilon(\pi^0\pi^0 J/\psi)\times      \nonumber \\
   & &({\mathcal B}(J/\psi\to e^+e^-)+{\mathcal B}(J/\psi\to \mu^+\mu^-))\Big ]
  \label{eq:obs_CS}
  \end{eqnarray}

The Born cross section is calculated with $\sigma_{\text{Born}} =\sigma_{\text{obs}}/[(1+\delta)(1+\delta^{\text{vac}})]$, where $(1+\delta)$ is a radiative correction factor obtained with KKMC~\cite{ref:kkmc}  and $(1+\delta^{\text{vac}})$ is a vacuum polarization factor following Ref.~\cite{ref:vac}. Note that due to initial state radiation to $e^+e^-$ resonant structures such as $Y(4260)$, $(1+\delta)$ depends on $E_{cm}$. The inputs and results are listed in Table~\ref{tab-xsecup}.  In cases where there is no statistically significant signal, the upper limits at $90\%$ confidence level are provided. For $N(Z_c^0)$  and $N(\pi^0\pi^0 J/\psi)$ the errors and upper limits are statistical only. A cap of 1 is set on R values. Figure~\ref{fig-rvse_bxsecvse}(a) and (b) show $R$ and $\sigma_{\text{Born}}$ as functions of ${E_{\rm cm}}$ with error bars that are statistical only.

\begin{table*}[htbp]
\begin{center}
\caption{Efficiencies, yields, $R=\frac{\sigma(e^+e^-\to\pi^0 Z_c(3900)^{0}\to\pi^0\pi^0 J/\psi)}{\sigma(e^+e^-\to\pi^0\pi^0 J/\psi)}$, and $\pi^0\pi^0 J/\psi$ Born cross sections at each energy point. For $N(Z_c^0)$  and $N(\pi^0\pi^0 J/\psi)$ errors and upper limits are statistical only. For $R$ and $\sigma_{\text{Born}}$, the first errors are statistical and second errors are systematic. The statistical uncertainties on the efficiencies
  are negligible. Upper limits of $R$ (90\% confidence level) include systematic errors. } \label{tab-xsecup}
\vspace{0.2cm}
\resizebox{!}{2.0cm}{
\begin{tabular}{|c|c|c|c|c|c|c|c|c|c|c|c|c|c|}
\hline 

${E_{\rm cm}} $ & $\mathcal L$ & $\epsilon(Z_c^0)$ & $\epsilon_{1}(\pi^0\pi^0 J/\psi)$ & $\epsilon_{2}(\pi^0\pi^0 J/\psi)$ & $\epsilon(\pi^0\pi^0 J/\psi)$ & $N(Z_c^0)$ & $N(\pi^0\pi^0 J/\psi)$ & $R$ & $1+\delta$ & $1+\delta^{vac}$ & $\sigma_{Born}$ (pb)\\
(GeV) & (pb$^{-1})$ & (\%) & (\%) & (\%) & (\%) & (90\% C. L.)  & & (90\% C. L.) & & & \\\hline

4.190 & 43.1 & $20.8$ & $20.4$ & $20.1$ & $20.2$ & $<11.1$ & $8.2\pm3.0$ & $0.71\pm0.45\pm0.04$ ($<1.00$) & 0.828 & 1.056 & $9.0\pm3.3\pm0.6$ \\\hline
4.210 & 54.6 & $21.5$ & $21.0$ & $20.8$ & $20.9$ & $<18.9$ & $26.6\pm5.4$ & $0.42\pm0.21\pm0.03$ ($<0.72$) & 0.813 & 1.057 & $22.7\pm4.6\pm1.5$ \\\hline
4.220 & 54.1 & $21.6$ & $21.2$ & $20.8$ & $21.1$ & $<12.6$ & $31.9\pm5.7$ & $0.18\pm0.14\pm0.02$ ($<0.41$) & 0.810 & 1.057 & $27.4\pm4.9\pm1.8$ \\\hline
4.230 & 1091.7 & $22.0$ & $21.1$ & $21.0$ & $21.0$ & $236.8\pm25.0$ & $825.1\pm29.8$ & $0.28\pm0.03\pm0.02$  & 0.805 & 1.056 & $35.4\pm1.3\pm2.2$ \\\hline
4.245 & 55.6 & $22.3$ & $21.6$ & $21.1$ & $21.5$ & $<15.2$ & $49.0\pm7.1$ & $0.15\pm0.10\pm0.02$ ($<0.32$)  & 0.806 & 1.056 & $40.3\pm5.8\pm2.7$ \\\hline
4.260 & 825.7 & $22.6$ & $21.2$ & $21.4$ & $21.2$ & $73.1\pm16.5$ & $507.3\pm23.4$ & $0.14\pm0.03\pm0.01$ & 0.815 & 1.054 & $28.3\pm1.3\pm1.8$ \\\hline
4.310 & 44.9 & $22.5$ & $20.4$ & $20.7$ & $20.5$ & $<7.9$ & $25.5\pm5.1$ & $0.07\pm0.12\pm0.01$ ($<0.29$)  & 0.916 & 1.052 & $24.1\pm4.9\pm1.6$ \\\hline
4.360 & 539.8 & $21.5$ & $18.8$ & $19.1$ & $18.9$ & $41.8\pm10.8$ & $182.8\pm14.2$ & $0.20\pm0.05\pm0.02$ & 1.038 & 1.051 & $13.8\pm1.1\pm0.9$ \\\hline
4.390 & 55.2 & $21.4$ & $17.7$ & $18.4$ & $17.7$ & $<5.2$ & $6.2\pm2.6$ & $0.00\pm1.02\pm0.00$ ($<0.71$) & 1.088 & 1.051 & $4.7\pm1.9\pm0.3$ \\\hline
4.420 & 44.7 & $21.7$ & $16.8$ & $17.9$ & $16.8$ & $<3.8$ & $2.9\pm2.1$ & $0.00\pm0.56\pm0.00$ ($<1.00$)  & 1.132 & 1.053 & $2.7\pm1.9\pm0.2$ \\\hline

\end{tabular}
}
\end{center}
\end{table*}

\begin{figure}[h]
\begin{center}
\includegraphics[width=9cm]{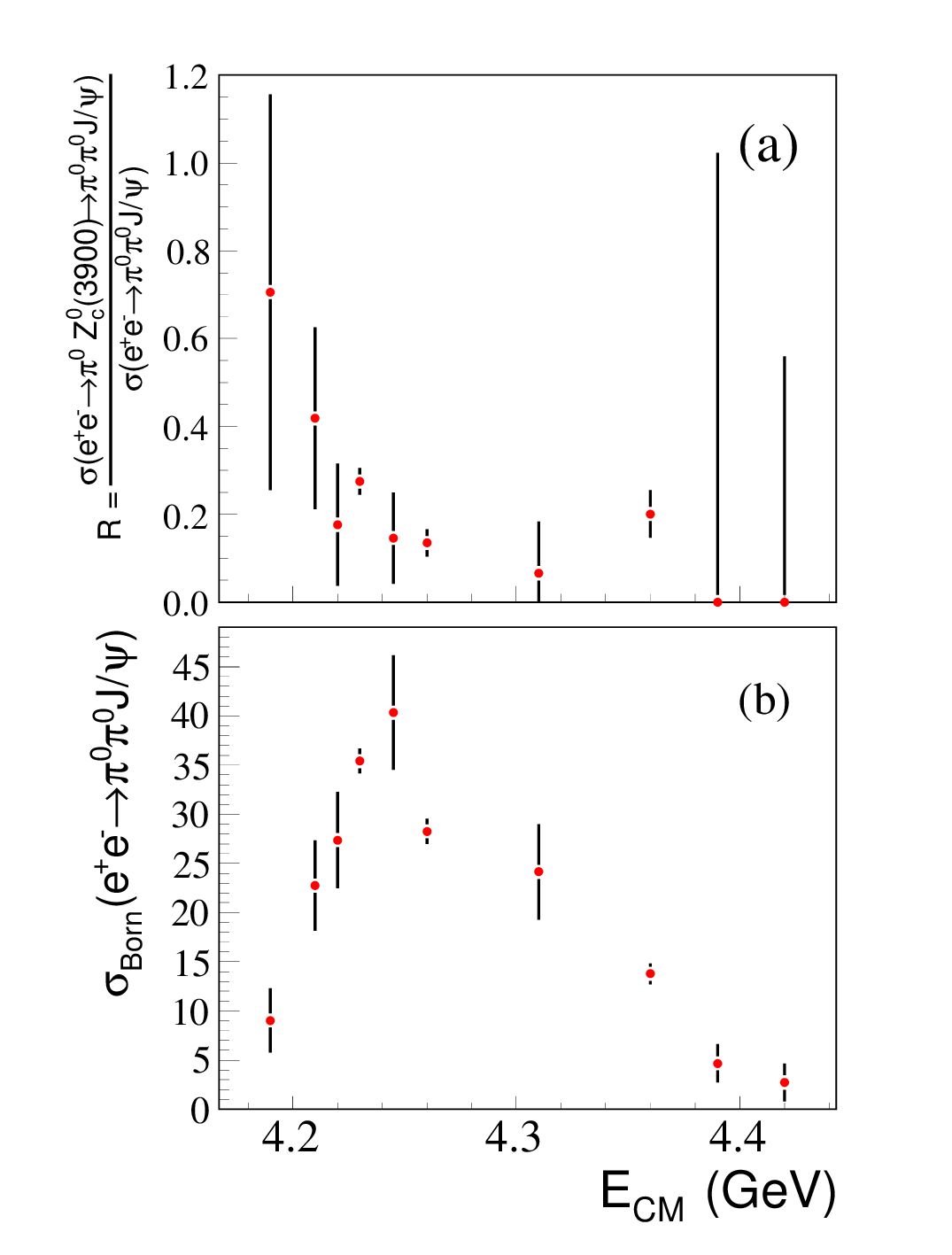}
\caption{(a) $R$ (see text) and (b) $\sigma_{Born}(e^+e^- \to \pi^0\pi^0 J/\psi$) as functions of ${E_{\rm cm}}$. Error bars are statistical only.} \label{fig-rvse_bxsecvse}
\end{center}
\end{figure}

We consider several sources of systematic uncertainty in the $Z_c(3900)^{0}$ mass and width measurements.  For the mass determination, the largest uncertainty is that associated with the absolute track momentum scale, estimated to be 2.0~MeV/$c^2$ based on the difference between the dilepton mass determined by the fit and the nominal $J/\psi$ mass. Uncertainty due to the knowledge of the beam energy is estimated to be 1.7~MeV/$c^2$ based on a study of $e^+e^- \to \mu^+ \mu^-$. Adjusting the cut on $\chi^2_{\rm 7C}$ by $\pm 30$ changes the mass by 1.2~MeV/$c^2$, which we assign as the systematic uncertainty associated with the kinematic fit.  To assess the uncertainty from the signal parameterization we change the phase-space factor from $pq$ to $p^3q^3$ (S-wave to P-wave) and find a 1.1~MeV/$c^2$ change in the mass.  Additional systematic effects associated with fitting-range dependence (0.8~MeV/$c^2$), background-shape sensitivity (0.3~MeV/$c^2$) and ${E_{\rm cm}}$ dependence (0.2~MeV/$c^2$) contribute at a lower level, leading to an overall systematic error in $M(Z_c(3900)^{0})$ of 3.2~MeV/$c^2$.  The measurement of $\Gamma(Z_c(3900)^{0})$ has a total systematic error of 8.2~MeV, which includes similarly sized contributions from the kinematic fitting procedure (4.6~MeV), background shape (4.1~MeV), fitting range (3.9~MeV), and ${E_{\rm cm}}$ (3.3~MeV), with a smaller effect due to the signal model (1.7~MeV) and none from the absolute mass scale. 

The uncertainties in $R$ and $\sigma_{Born}$ include contributions from the luminosity (0\% for $R$ and 1.0\% for $\sigma_{Born}$)~\cite{Ablikim:2015nan}, tracking efficiency (0\% and 2.0\%)~\cite{Ablikim:2011kv},  $\pi^0$ selection efficiency (0\% and 4.0\%)~\cite{Ablikim:2010zn},   muon identification  efficiency (0\% and 3.0\%), background shape (3.0\% and 0.6\%),  peaking backgrounds (1.4\% and 1.4\%),   fitting range  (2.6\% and 0.6\%),   kinematic fit (2.2\% and 1.7\%),   intermediate-state branching ratios (0\% and 0.5\%),   signal parameterization (1.9\% and 1.9\%),  input cross section line shape in KKMC (0\% and 0.6\%) \cite{Yuan:2007sj,Lees:2012cn}, line shape of $e^+e^-\to\pi^0Z_c^0$ (1.1\% to 12.3\% and 0\% to 3.2\%, depending on $E_{\text{cm}}$), and decay models of $\pi^0\pi^0 J/\psi$ in the MC (0.2\% to 6.3\% and 0.2\% to 6.3\%).  An uncertainty of $0\%$ in $R$ signifies that the effect of that source of systematic uncertainty cancels in the ratio.   Results for $R$ and $\sigma_{Born}$ with systematic errors are given in Table~\ref{tab-xsecup}. In cases where there is no statistically significant signal, upper limits are defined as sums of $90\%$ confidence level statistical upper limits plus systematic errors.

In summary, we have observed a new charmonium-like state $Z_c(3900)^{0}$ in $e^+e^-\to \pi^0\pi^0 J/\psi$ with a statistical significance of 10.4$\sigma$. The mass and width of $Z_c(3900)^{0}$ are measured to be $3894.8\pm2.3\pm3.2$ MeV/$c^2$  and $29.6\pm8.2\pm8.2$~MeV, respectively.  We interpret this state as the neutral partner of the four-quark state candidate $Z_c(3900)^\pm$,  since it decays to $\pi^0 J/\psi$ and its mass is close to the mass of $Z_c(3900)^\pm$. The previous report of 3.5$\sigma$ evidence  for $Z_c(3900)^{0}$~\cite{Xiao:2013iha} included values of the mass and width that are consistent with our results, but are much less precise.  We have also measured the cross section ratio $R=\frac{\sigma(e^+e^-\to\pi^0 Z_c(3900)^{0}\to\pi^0\pi^0 J/\psi)}{\sigma(e^+e^-\to\pi^0\pi^0 J/\psi)}$ and the Born cross section for $e^+e^-\to\pi^0\pi^0 J/\psi$ in the energy range from $4.190$ to $4.420$~GeV.  The measured Born cross sections are about half of those for $e^+e^-\to\pi^+\pi^- J/\psi$ that were measured by Belle \cite{Liu:2013dau} , consistent with the isospin symmetry expectation for resonances. 

The BESIII collaboration thanks the staff of BEPCII and the IHEP
computing center for their strong support. This work is supported in
part by National Key Basic Research Program of China under Contract
No.~2015CB856700; National Natural Science Foundation of China (NSFC)
under Contracts Nos.~11125525, 11235011, 11322544, 11335008, 11425524;
the Chinese Academy of Sciences (CAS) Large-Scale Scientific Facility
Program; the CAS Center for Excellence in Particle Physics (CCEPP);
the Collaborative Innovation Center for Particles and Interactions
(CICPI); Joint Large-Scale Scientific Facility Funds of the NSFC and
CAS under Contracts Nos.~11179007, U1232201, U1332201; CAS under
Contracts Nos.~KJCX2-YW-N29, KJCX2-YW-N45; 100 Talents Program of CAS;
INPAC and Shanghai Key Laboratory for Particle Physics and Cosmology;
German Research Foundation DFG under Contract No.~Collaborative
Research Center CRC-1044; Istituto Nazionale di Fisica Nucleare,
Italy; Ministry of Development of Turkey under Contract
No.~DPT2006K-120470; Russian Foundation for Basic Research under
Contract No.~14-07-91152; U.S.~Department of Energy under Contracts
Nos.~DE-FG02-04ER41291, DE-FG02-05ER41374, DE-FG02-94ER40823,
DESC0010118; U.S.~National Science Foundation; University of Groningen
(RuG) and the Helmholtzzentrum fuer Schwerionenforschung GmbH (GSI),
Darmstadt; WCU Program of National Research Foundation of Korea under
Contract No.~R32-2008-000-10155-0.


\begin{thebibliography}{99}
\bibitem{ref:ecm} 
$E_{cm}$ values quoted in this Letter are nominal values based on the BEPCII accelerator control system, and true center-of-mass energies are lower by 2-3~MeV.  This difference has a minimal effect on the analysis reported here and is treated as a systematic uncertainty.
\end{thebibliography}
\end{document}